# GAMMA CASCADE TRANSITION OF $^{51}$V(n$_{th}$,γ)$^{52}$V REACTION

P. D. Khang[1], N. X. Hai[1], H. H. Thang[1], V. H. Tan[2]
N. A. Son[3], N. D. Hoa[3], D. M. Trinh[3]

[1]*Vietnam Atomic Energy Institute, 59 Ly Thuong Kiet, Hanoi, Vietnam*
[2]*Vietnam Agency for Radiation and Nuclear Safety, 113 Tran Duy Hung, Hanoi, Vietnam*
[3]*University of Dalat, 01 Phu Dong Thien Vuong, Dalat, Vietnam*

**Abstract:** The thermal neutron capture gamma radiations for $^{51}$V(n, γ)$^{52}$V reaction have been studied at Dalat Nuclear Research Reactor (DNRR). The gamma two-step cascade transition was measured by event-event coincidence spectrometer. The added-neutron binding energy in $^{52}$V was measured as 7.31 MeV. Energy and intensity transition of cascades were consistent with prediction of single particle model. Further more, the spin and parity of levels were confined.

## 1. Introduction

The exited states of $^{52}$V have been studied from 1960s, the $^{52}$V level scheme has mainly been performed by means of the $^{51}$V(d, p)$^{52}$V and $^{51}$V(n, γ)$^{52}$V reactions. The energy and intensity of proton groups in the (d, p) reaction on $^{51}$V for states in $^{52}$V up to 3.3 MeV were measured [[1][2][3][4]]. The (d, p) reaction studies were in good agreement with the analysis of the $^{51}$V(n, γ)$^{52}$V reaction, investigated by single gamma spectra or fast-coincidence gamma spectrometry [1][2][3][6].

The Ritz-combination principle was used to construct nuclear level scheme from low energy neutron capture gamma ray. The bent crystal spectrometer was reduced the interference from fortuitous combinations. This technique can be applied for gamma ray energies between 30 keV and several hundred keV. Above 1 MeV, the efficiency and energy resolution of bent crystal spectrometer is decrease, and other experimental techniques should be used [2][5].

Several studies of the gamma spectrum arising from thermal neutron capture in the $^{51}$V(n, γ)$^{52}$V reaction have been carried out. The spectrum above 3 MeV was determined using a pair spectrometer. Groshev et al. detected gammas down to 0.4 MeV using a Compton spectrometer, but with relatively poor resolution at lower end. The low-lying region has been investigated mostly by interpreting singles spectra from NaI(Tl) and HPGe spectrometers [2]. In Ref. [1], the many levels of low energy were found. In this, coincidence and sum-coincidence measurement techniques were used. The sum-coincidence method for capture-gamma work is based upon the simple fact that in any nuclear decay scheme caused by gamma de-excitation from a thermal neutron capture state to the ground state, the energy of gamma-rays related in cascade must sum to the neutron binding energy. S. Michaelsen et al studied the level schemes of $^{52}$V with high sensitivity and very good energy resolution, via thermal neutron capture reactions. In $^{52}$V, he had mainly focused on states between 3.5 and 7 MeV excitation. The populating and decay radiations of most of these states have been established. The density of states in spin and parity windows are almost completed up to half the binding energies [7] .

The nuclear data evaluation of $^{52}$V isotope in library was shown the lack of quantum properties of energy levels and the not singular value of spin of some energy levels. The adopted data set looks fairly complete, either missing or being off by about 3%. Energies are probably good to about 0.1 keV. It is hard to assign uncertainties to the intensities, as those of Michaelsen et al. were reported as about 10% but the agreements of the sums into and out off various levels are usually better than 10% [8]. Therefore, the $^{52}$V need to be more experiment information in order to study of nuclear structure.

The nucleus $^{52}$V has three protons and one neutron outside a closed shell core having the structure of the doubly magic $^{48}$Ca. It is in a region in which shell model calculations involving coupling between extra-core nucleons in different shell are known to be particularly appropriate [2][1].

Today, with development of HPGe detector, the sum-coincidence measurement method has been developed to "event-event" coincidence method which is digitalized. In this experiment, the gamma intensities, excited levels of $^{52}$V in energy region below separation neutron energy were measured by "event-event" coincidence method. The experimental results were compared with prediction of single particle model.

## 2. Experiment
### 2.1. The experimental arrangement

The experimental system has been installed at the tangential beam port of the DNRR. The thermal neutron beam was filtered by S, Pb and Si. The neutron flux, the cadmium ratio was 900 (1 cm thinness of cadmium) and the neutron beam at the target position was $1.02 \times 10^6$ n.cm$^{-2}$s$^{-1}$. The collimator was made a mixture of paraffin and boron. The distance from the endcap of detectors to the neutron beam center was 4 cm; lead bricks of 10 cm thickness were used to surround the detectors as gamma shields which the background count rate in the 0.2 ÷ 8 MeV range was guaranteed less than 400 counts per second (cps). Two plates of 2 mm thick lead were placed between the detectors and target to decrease the number of backscattered gamma rays and to filter out X-ray.

The electronics configuration was used in those gamma-gamma coincidence experiments are shown in Fig. 1, the parameters of system were setup by method in Ref. [9]. The detector signals are amplified with 572 amplifier (AMP) modules with a shaping time of 3.0 μs and about 1 keV/1 channel. Output signals of the amplifiers are digitized by 7072 analog-to-digital converter (ADC) modules. The timing signals of both detectors are put through 474 timing filter amplifier (TFA) modules. The shaped and amplified timing signals by 474 TFA are plugged into 584 CFD modules, which are used in slow rise time rejection option (SRT) mode. The CFD output signal of the first channel (using GPC20 detector) is used as 556 time-to-amplitude converter (TAC) start signal. The CFD output signal of the second channel (using GC2018 detector) is delayed 100 ns and served as a TAC stop signal. The full scale of TAC is set at 500 ns, and output signal is digitized in 8713 ADC with selection of 1024 channels for a 10 V input pulse. The TAC ''Valid Convert'' signal is used to gate 7072 ADC, and the delay for synchronizing with AMP output signal is implemented by interface software. Recorded coincident events have three values, including coincidence gamma-ray energies from detector 1, detector 2 and time interval between two γ-rays in one detection of pair event.

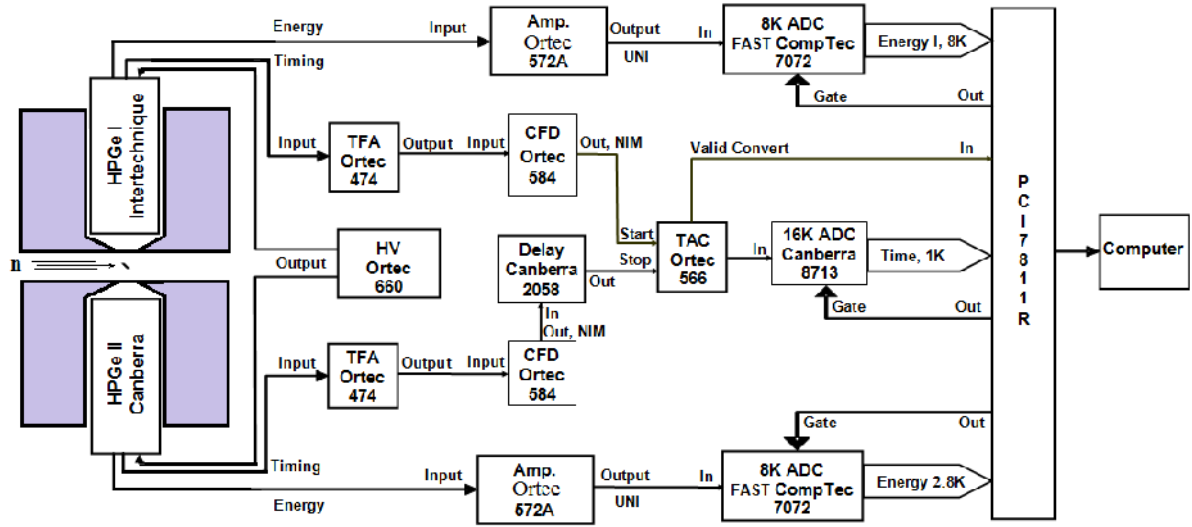

**Fig 1.** *The electronics configuration.*

The target is natural Vanadium which the rich of $^{51}$V is 99.75%, the thermal neutron capture cross section of $^{51}$V is σ = 4.9 barn [10]. The target was placed between two detectors. Experimental data collected 280 hours total. The coincidence count rate was about 34 cps. The programming interface was set up in event-event coincidence mode. The data on the deposited gamma ray energies and time intervals were recoded and processed off-line by summation of amplitude method.

**2.2. Data analysis**

From the recorded information a set of gamma rays spectra were obtained that belong to all tow step cascades that end at preselected final levels in $^{52}$V. We call these spectra two-step gamma cascades (TSC) spectra and the corresponding TSC final level. Each of these spectra were constructed from deposited energy in one of the detectors under condition that the gamma ray energy sum from both detectors fell within the region of full energy line corresponding to a preselected level. The energies and $J^\pi$ of these final levels are listed in Table 1. Note, that in one case two of the levels were not resolved.

While constructing the TSC spectra, the background due to accidental coincidences and Compton scattering was subtracted: Compton background was subtracted by choosing background regions on two sides of peak in the spectrum of energy sums; and choosing time windows, selecting three intervals of detection-time difference, was adjusted to isolate the net signal from the background due to accidental coincidences.

**Table 1.** *The energies and $J^\pi$ of these final levels.*

| $N^0$ | Final level (keV) | Spin ($J^\pi$) | $N^o$ | Final level (keV) | Spin ($J^\pi$) |
|---|---|---|---|---|---|
| *1* | 0.00 | $3^+$ | *4* | 147.85 | $4^+$ |
| *2* | 17.16 | $2^+$ | *5* | 436.30 | $2^+, 3^+$ |
| *3* | 22.29 | $5^+$ | *6* | 793.34 | $2^+, 3^+$ |

The constructed TSC spectra were corrected for the energy dependence of the full energy line efficiencies of both detectors. In addition, corrections for the vetoing effects caused by the detection of gamma rays following the decay of the TSC final level and for the effects of gamma-gamma angular correlations were applied. After these corrections the TSC spectra were converted into spectra expressed in absolute TSC intensities. This conversion requires knowledge of the TSC intensity of at least one TSC and it was performed with data from library.

The Fig. 2 is a part of spectrum of gamma ray energy sums accumulated from the event-event mode data obtained from the coincidence neutron capture $^{51}$V. The Fig. 3 is TSC spectrum of $^{52}$V belong to final level at 0 and 18 keV.

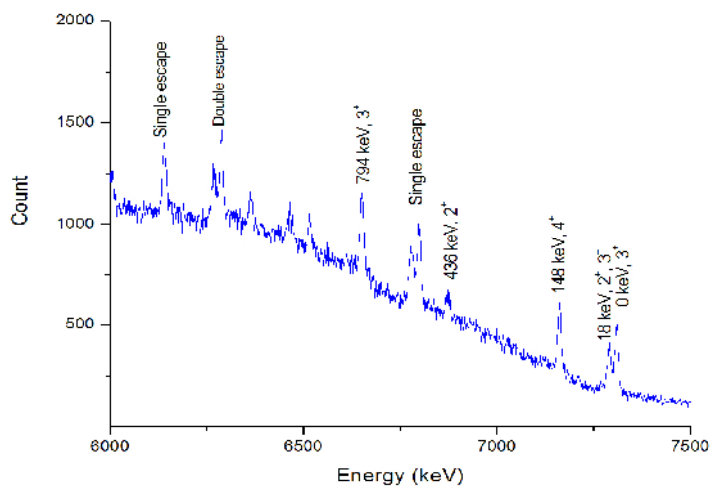

**Fig 2.** *A part in gamma ray energy sums of spectrum.*

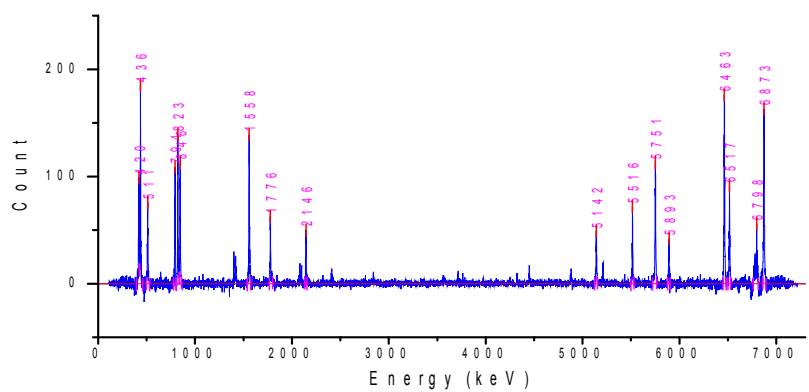

**Fig 3.** *TSC spectrum of $^{52}$V belong to final level at 0 and 18 keV.*

## 2.3. Spin assignment and theoretical investigation of the $^{52}$V levels

In neutron capture reaction, if J is spin of target then J ± ½ are probability spins of compound nuclear. The emitted gamma rays usually have E1, M1, E2 or M1+E2 called the multi-polarity orders of the radiation. The multi-polarity orders of the radiation are determined:

$$|J_i - J_f| \leq L \leq J_i + J_f \qquad (1)$$

L = 1 is electric dipole (E1) and magnetic dipole (M1), L = 2 is electric quadruple (E2) and magnetic quadruple (M2). The parity will be $(-1)^L$ for electric transition, the parity will be $(-1)^{L+1}$ for magnetic transition [11].

According to the single particle model, the -ray transition probability is predicted by Eq. 2 [12].

$$T^{EL} = \frac{8}{L[(2L+1)!!]^2 \hbar} \cdot 2^L \left(\frac{E}{\hbar c}\right)^{2L+1} B(EL)\downarrow$$

$$T^{ML} = \frac{8}{L[(2L+1)!!]^2 \hbar} \cdot \mu_N^2 \cdot 2^{L-1} \left(\frac{E}{\hbar c}\right)^{2L+1} B(ML)\downarrow \quad (2)$$

where: $\hbar c = 197.327 \times 10^{-10}$ keV.cm, $\hbar = 6.58212 \times 10^{-19}$ keV.s, $e^2 = 1.440 \times 10^{-10}$ keV.cm, $\mu_N^2 = 1.5922$ $^{-23}$ $^3$ $^{-24}$ $^2$

Both $B(EL)\downarrow$ and $B(ML)\downarrow$ are the reduced downward probabilities for electric and magnetic transition respectively. Using the Weisskopf derived for the single particle estimates for these matrix elements based on the shell model in Eq. 3 [12]:

$$B(EL) = \frac{1}{4\pi} \left(\frac{3}{L}\right)^2 R^{2L}$$

$$B(ML) = \frac{10}{\pi b^{L-1}} \left(\frac{3}{3+L}\right)^2 R^{2L-2} \quad R = 1.2 \times 10^{-13} A^{1/3} \text{cm}. \quad (3)$$

From Eq. 2 and Eq. 3, the transition probabilities of $^{52}$V can be calculated by Eq. 4.

$$T^{E1} = 1.4280 \times 10^6 E^3$$
$$T^{E2} = 1.4127 \times 10^{-5} E^5 \quad (4)$$
$$T^{M1} = 3.1483 \times 10^4 E^3$$
$$T^{M2} = 3.1143 \times 10^{-7} E^5$$

The predicted gamma transition according to Eq. 4 for $^{52}$V are shown in table 3.

### 3. Result and discussion

**3.1. The gamma two-step cascade energies, intensities and intermediate levels**

By this work, the 36 pairs of gamma two-step cascades were recorded, the results are shown in table 2.

**Table 2.** *The gamma two-step cascade energies and intensities of $^{52}V$ in $^{51}V(n_{th},\gamma)^{52}V$ reaction.*

| N° | $E_1$ (keV) | $E_2$ (keV) | $E_L$ (keV) | $I_{\gamma-\gamma}$ (%) |
|---|---|---|---|---|
| | $E_1+E_2 = 7310.68$ keV, $E_f = 0$ keV | | | |
| 1 | 6875.09(102) | 436.30(52) | 436.34 | 2.913(24) |
| 2 | 6518.05(94) | 793.34(62) | 793.34 | 3.700(27) |
| 3 | 6465.04(98) | 845.35(64) | 845.64 | 4.324(45) |
| 4 | 5892.97(142) | 1418.42(77) | 1418.42 | 2.511(22) |
| 5 | 5752.96(123) | 1558.44(88) | 1558.44 | 9.461(67) |
| 6 | 5578.93(104) | 1732.46(89) | 1732.46 | 0.487(65) |
| 7 | 5516.93(76) | 1795.47(92) | 1795.47 | 0.855(11) |
| 8 | 5211.89(89) | 2101.51(114) | 2101.51 | 0.974(23) |
| 9 | 5142.88(98) | 2169.51(121) | 2167.80 | 0.540(14) |
| 10 | 4993.86(102) | 2317.53(124) | 2316.82 | 0.294(12) |
| 11 | 4884.85(114) | 2427.55(146) | 2427.55 | 0.435(09) |
| $E_1 + E_2$ = 7293.52 keV, $E_f$ = 17.16 keV, 22.29 keV | | | | |
| 12 | 6875.09(102) | 419.30(52) | 436.34 | 1.849(24) |
| 13 | 6465.04(98) | 823.35(63) | 845.64 | 4.433(34) |
| 14 | 5892.97(142) | 1401.42(77) | 1418.42 | 1.317(26) |
| 15 | 5516.93(76) | 1778.47(89) | 1795.47 | 3.776(27) |
| 16 | 5211.89(89) | 2083.50(112) | 2098.79 | 0.837(11) |
| 17 | 5142.88(98) | 2146.51(121) | 2167.80 | 3.382(32) |
| 18 | 4884.85(114) | 2410.54(132) | 2427.55 | 0.802(14) |
| 19 | 4452.80(146) | 2842.60(145) | 2857.88 | 0.871(17) |
| 20 | 3579.69(165) | 3716.71(168) | 3730.99 | 0.625(13) |
| $E_1 + E_2$ = 7162.83 keV, $E_f$ = 147.85 keV | | | | |
| 21 | 6875.09(102) | 295.28(49) | 436.34 | 1.130(22) |
| 22 | 6518.05(94) | 645.33(60) | 792.63 | 7.957(58) |
| 23 | 6465.04(98) | 698.33(59) | 845.64 | 1.229(23) |
| 24 | 5752.96(123) | 1410.42(78) | 1557.72 | 1.569(64) |
| 25 | 5551.93(68) | 1612.45(54) | 1758.75 | 0.732(20) |
| 26 | 5211.89(89) | 1953.49(95) | 2098.79 | 1.780((54) |
| 27 | 5142.88(98) | 2021.50(102) | 2167.80 | 0.949((43) |
| 28 | 4452.80(146) | 2710.58(165) | 2857.88 | 0.581(24) |
| $E_1 + E_2$ = 6874.51 keV, $E_f$ = 436.34 keV | | | | |
| 29 | 5892.97(142) | 982.37(66) | 1417.71 | 0.539(44) |
| 30 | 5516.93(76) | 1358.41(73) | 1793.75 | 4.012(68) |
| 31 | 5211.89(89) | 1664.45(54) | 2098.79 | 1.669(51) |
| $E_1 + E_2$ = 6517.34 keV, $E_f$ = 793.34 keV | | | | |
| 32 | 5516.93(76) | 1002.37(70) | 1793.75 | 1.654(60) |
| 33 | 5211.89(89) | 1307.41(72) | 2098.79 | 1.262(51) |
| 34 | 4884.85(114) | 1634.45(56) | 2425.83 | 0.954(30) |
| $E_1 + E_2$ = 1793.38 keV, $E_f$ = 5516.62 keV | | | | |
| 35 | 1358.4(73) | 436.30(55) | 436.34 | 0.018(14) |
| 36 | 1002.37(70) | 793.34(62) | 793.63 | 0.025(12) |

***Note:*** *$E_1$ (keV) is energy of the primary gamma transition, $E_2$ (keV) is energy of secondary gamma transition, $E_L$ (keV) is the intermediate level of cascade and $I_{\gamma\gamma}$ (%) is relative intensity of gamma cascade transition.*

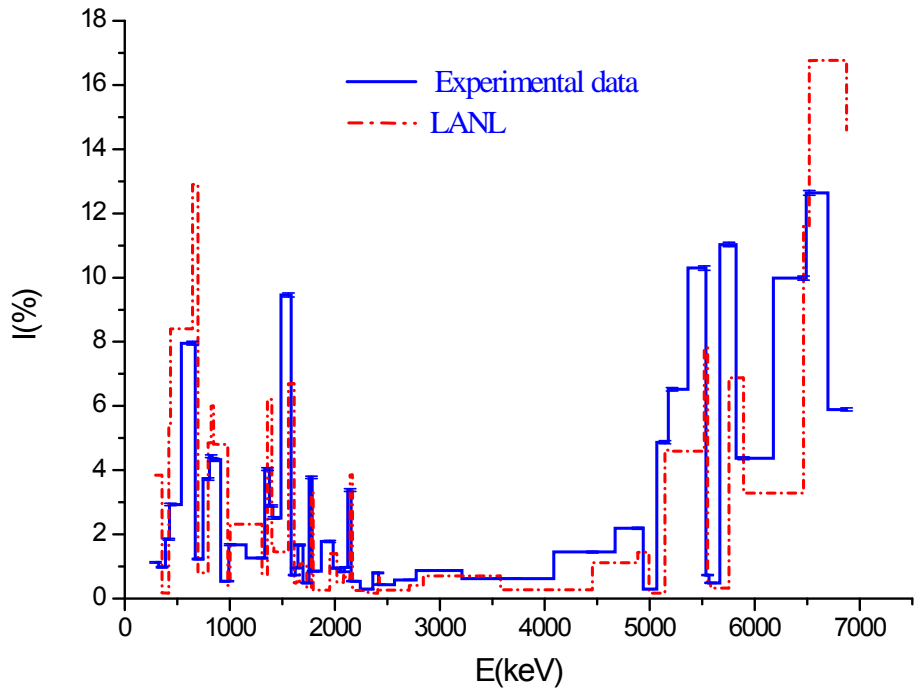

**Fig. 4.** *The distribution of gamma transition intensities to energy.*

The Fig 4. is the distribution of gamma transition intensities to energy. The line and dot are this work and in Ref. 8 respectively.

### 3.2. Parities, spin assignments and transition probabilities

According to the shell model of nucleus, the structure of $^{51}$V is $1s_{1/2}^4 1p_{3/2}^8 1p_{1/2}^4 1d_{5/2}^{12} 2s_{1/2}^4 1d_{3/2}^8 1f_{7/2}^{11}$, so the j of $^{51}$V in ground state is $7/2^-$. The j of the state formed on neutron capture by $^{51}$V may have the value of $3^-$ or $4^-$ (an s-wave neutron can be captured). The j of $^{52}$V in ground state is $3^+$[1][2][3][4]. The parity of the $^{52}$V excited state is opposite to the ground state, and the directly gamma emitted compound to ground sate, so transition between these states (transition 1$^{st}$ order) must be of the E1 type.

Further more, refer to the word of Schwager [1], the possibilities j of $^{52}$V* in intermediate state are $0^+$, $1^+$, $2^+$, $3^+$, $4^+$, $5^+$, $6^+$, $7^+$, $8^+$ and $9^+$. According to single particle model nuclear structure, applying condition (1), and equation (4), the possibilities spins, parities of levels and transition probabilities of $^{52}$V exciting in thermal neutron capture reaction, are shown in table 3.

**Table 3.** *The spins, parities of levels, and transition probabilities of $^{52}$V exciting in thermal neutron capture reaction.*

| Level (keV) | E (keV) | $j_f^\pi$ (Exp) | $j_f^\pi$ (Ref. [2][3][4][5][6][7]) | $T^{M1}$ (Exp.) | $T^{M1}$ (Model) | $(T^{M1} - T^{M1})/T^{M1}$ |
|---|---|---|---|---|---|---|
| ($2^+$, $3^+$) 436.30 | 436.30 | $3^+$ | $3^+$ | 0.494 | 0.455 | 0,09 |

| | 419.30 | $2^+$ | $2^+$ | 0.314 | 0.404 | -0,22 |
| --- | --- | --- | --- | --- | --- | --- |
| | 295.28 | $1^+$ | $1^+$ | 0.192 | 0.141 | 0,36 |
| ($2^+$, $3^+$) 793.34 | 793.34 | $3^+$ | $3^+$ | 0.293 | 0.614 | -0,52 |
| | 645.33 | $4^+$ | $4^+$ | 0.630 | 0.330 | 0,91 |
| | 356.29 | $2^+$, $3^+$ | $2^+$ | 0.077 | 0.056 | 0,38 |
| ($3^+$, $4^+$) 845.35 | 845.35 | $3^+$ | $3^+$ | 0.433 | 0.402 | 0,08 |
| | 823.35 | $5^+$ | $5^+$ | 0.444 | 0.371 | 0,20 |
| | 698.33 | $4^+$ | $4^+$ | 0.123 | 0.227 | -0,46 |
| ($2^+$, $3^+$) 1417.71 | 1418.42 | $3^+$ | $3^+$ | 0.575 | 0.435 | 0,32 |
| | 1401.42 | $2^+$ | $2^+$ | 0.302 | 0.420 | -0,28 |
| | 982.37 | $2^+$, $3^+$ | $2^+$ | 0.123 | 0.145 | -0,15 |
| ($3^+$, $4^+$) 1557.72 | 1558.44 | $3^+$ | $3^+$ | 0.858 | 0.574 | 0,49 |
| | 1410.42 | $4^+$ | $4^+$ | 0.142 | 0.425 | -0,67 |
| ($2^+$, $3^+$) 1793.75 | 1795.47 | $3^+$ | $3^+$ | 0.083 | 0.387 | -0,79 |
| | 1778.47 | $2^+$ | $2^+$ | 0.367 | 0.376 | -0,02 |
| | 1358.41 | $2^+$, $3^+$ | $2^+$ | 0.390 | 0.168 | 1,32 |
| | 1002.37 | $2^+$, $3^+$ | $3^+$ | 0.161 | 0.067 | 1,40 |
| ($2^+$, $3^+$) 2101.51 | 2101.51 | $3^+$ | $3^+$ | 0.149 | 0.284 | -0,48 |
| | 2083.50 | $2^+$ | $2^+$ | 0.128 | 0.277 | -0,54 |
| | 1953.49 | $4^+$ | $4^+$ | 0.273 | 0.228 | 0,20 |
| | 1664.45 | $2^+$, $3^+$ | $2^+$ | 0.256 | 0.141 | 0,82 |
| | 1307.41 | $2^+$, $3^+$ | $3^+$ | 0.193 | 0.068 | 1,84 |
| ($3^+$, $4^+$) 2169.51 | 2169.51 | $3^+$ | $3^+$ | 0.111 | 0.359 | -0,69 |
| | 2146.51 | $5^+$ | $5^+$ | 0.694 | 0.348 | 0,99 |
| | 2021.50 | $4^+$ | $4^+$ | 0.195 | 0.291 | -0,33 |
| ($2^+$, $3^+$) 2427.55 | 2427.55 | $3^+$ | $3^+$ | 0.199 | 0.437 | -0,54 |
| | 2410.54 | $2^+$ | $2^+$ | 0.366 | 0.428 | -0,14 |
| | 1634.45 | $2^+$, $3^+$ | $3^+$ | 0.435 | 0.133 | 2,27 |
| ($3^+$, $4^+$) 2857.88 | 2842.60 | $2^+$ | $2^+$ | 0.600 | 0.534 | 0,12 |
| | 2710.58 | $4^+$ | $4^+$ | 0.400 | 0.463 | -0,14 |

## 4. Discussion

In this work the final state 17 keV and 22 keV appeared in one peak of the summation spectrum at 18 keV. It is not solved in summation spectrum but in TSC corresponding this peak, the gamma ray energies are separated in two groups, one has summation energies of 7.31 - 0.17 MeV and other has summation energy of 7.31 - 0.22 MeV.

The neutron binding energy of $^{52}$V appeared in summation spectrum at 7.31 MeV that is consisted with previous research. The spin of gamma ray energies of 356.29 keV, 1002.37 keV, 1307.41 keV, 1358.41 keV, 1634.45 keV and 1664.45 keV are not singular and can be accepted $2^+$ and $3^+$.

The transition probabilities of $^{52}$V are rather consisted with the prediction of single particle model. The difference of experimental and calculation results are about 2 to 22%, it is in range of experiment errors.

## 5. Conclusion

In this work, the 36 pairs of gamma two-step cascades were recorded, the results are consisted with prediction of single particle model. The event-event coincidence method is very useful for research of gamma cascade transition which can separated unresolved levels by detectors.